\documentclass[twocolumn,showpacs,preprintnumbers,amsmath,amssymb]{revtex4}
\begin{document}
\draft
\title{
$D_s(2317)$ and $D_s(2457)$ from HQET Sum Rules
}
\author{
Yuan-Ben Dai$^{\:a}$, Chao-Shang Huang$^{\:a}$, Chun Liu$^{\:a}$,
and Shi-Lin Zhu$^{\:b}$
}
\affiliation{
$^a$Institute of Theoretical Physics, Chinese Academy of Sciences,
PO Box 2735, Beijing 100080, China\\
$^b$Department of Physics, Peking University, Beijing 100871, China
}

\thispagestyle{empty}
\setcounter{page}{1}

\begin{abstract}
Within the framework of the heavy quark effective theory, QCD sum
rules are used to calculate the masses of p-wave $\bar c s$
states.  The results for $0^+$ and $1^+$ states with the angular
momentum of the light component $j_l=1/2$ are consistent with the
experimental values for $D_s(2317)$ and $D_s(2457)$, respectively.
\end{abstract}


\newpage

\maketitle

\section{Introduction}
\label{sec:introduction}

By the end of April, 2003, BaBar Collaboration announced the discovery of
a $D_s$ resonance state, $D_s(2317)$ with a narrow width \cite{babar}.
Its mass is surprisingly small compared to quark model expectations.
Lately CLEO \cite{cleo} confirmed it and announced a new state $D_s(2457)$
which also appears in BaBar's data.  BELLE confirmed both of the above
results \cite{belle}.

The quantum numbers and decay modes of these two states are given in
the following table.
\begin{table}
\label{table1}
\caption{
Spin-parity numbers and decay modes
}
\vskip 5mm
\begin{center}
\begin{tabular}{ccc}
\hline
       & $D_s(2317)$ & $D_s(2457)$ \\ \hline
$J^P$  & $0^+$       & $1^+$       \\
Decays & $D_s\pi^0$  & $D_s^*\pi^0$\\ \hline
\end{tabular}
\end{center}
\end{table}

The experimental discovery of these two states has drawn a lot of
theoretical attentions.  The narrowness of these states is
understood as follows. Their low masses forbid them to decay to
$D^{(\ast)} K$. For $D_s(2317)$ the next possible decay mode $D_s
\eta$ can not open either.  But the decay mode $D_s \pi^0$ is
viable due to the small mixing between $\eta$ and $\pi^0$ through
isospin-violation. The key point is to understand their low
masses. Previously quark models estimated the $0^+$ state to be
over $100$ MeV higher than $D_s(2317)$ \cite{qm}.  The model using
the heavy quark mass expansion of the relativistic Bethe-Salpeter
equation in \cite{jin} predicted a lower value $2.369$ GeV of
$0^+$ mass which is still 50 MeV higher than the experimental
data.  They were interpreted as the parity conjugate states of the
$D_s(0^-,1^-)$ mesons in the framework of chiral symmetry
\cite{chiral}.  A four-quark state explanation to $D_s(2317)$ was
suggested in Ref. \cite{cheng}.  Ref. \cite{van} argued that the
low mass of $D_s(2317)$ could arise from the mixing between $D K$
continuum and the lowest scalar nonet.  For other discussions, see
{\it e.g.} Refs. \cite{other}.

In this talk we report our calculations of the masses of p-wave $D_s$
states from heavy quark effective theory (HQET) sum rules \cite{ours}.
S-wave, non-strange heavy mesons have been studied with HQET sum rules,
{\it e.g.} in \cite{neubert}, and the strange ones in \cite{luo}.  The
p-wave, non-strange heavy mesons have been studied up to the order of
${\cal O}(1/m_Q)$ \cite{dai,daizhu,colangelo}.  They were also
analyzed in full QCD sum rules \cite{old}.

\section{HQET Description}
\label{sec:hqet}

Let us first describe the p-wave $D_s$ states.  There
are four states: $D_{s0}^*(0^+)$, $D_{s1}^*(1^+)$, $D_{s1}(1^+)$ and
$D_{s2}^*(2^+)$.  These kinds of heavy mesons can be systematically studied
by HQET \cite{hqet}.  In the heavy quark limit,
$m_Q/\Lambda_{\rm QCD}\to \infty$, there is heavy quark spin-flavor
symmetry.  After the redefinition of the quark field,
$P_+Q(x)=\exp (-im_Qv\cdot x)h_v(x)$, where
$P_+=\frac{1}{2}(1+\not\! v)$, the HQET Lagrangian for heavy quark is
simplified to be
\begin{equation}
{\cal L}_{\rm HQET} = \bar{h}_v i v \! \cdot \! D\, h_v\,.
\label{hqet}
\end{equation}
in this limit. The heavy quark spin symmetry implies that the
angular momentum of the light quark system $j_l$ is a good quantum
number. The four p-wave heavy mesons can be grouped into two
doublets ($D_{s0}^*$, $D_{s1}^*$) and ($D_{s1}$, $D_{s2}^*$).
$j_l=1/2$ for ($D_{s0}^*, D_{s1}^*$) and  $j_l=3/2$ for ($D_{s1},
D_{s2}^*$).  It is important to note that although $D_{s1}^*$ and
$D_{s1}$ have same total quantum numbers, they are clearly
separated in the heavy quark limit.  Their mixing is at the order
of $1/m_c$, which can be studied in HQET. In HQET, the meson
masses are expanded as
\begin{equation}
M=m_Q + \bar{\Lambda}_s \,,
\label{mass}
\end{equation}
where $\bar{\Lambda}_s$ are independent of heavy quark flavors in
the heavy quark limit.

To study properties of hadrons with QCD sum rules , the
interpolating currents carrying the same quantum numbers in the
HQET as those of the hadrons should be used. For the p-wave heavy
mesons, we write the currents as follows \cite{dai},
\begin{equation}
J=\frac{1}{\sqrt{2}}\bar{h}_v\Gamma D_{\rho} s\,,
\label{current}
\end{equation}
with $\Gamma$ denoting some $\gamma$ matrices, $\Gamma$ is found to be
\begin{equation}
\begin{array}{llll}
\Gamma&=&-\gamma_t^{\rho} ~~~&{\rm for}~~~ D_{s0}^*\,,\\
\Gamma&=&\gamma_5\gamma_t^{\mu}\gamma_t^{\rho} ~~~&
{\rm for}~~~ D_{s1}^*\,,\\
\Gamma&=&\displaystyle -\sqrt{\frac{3}{2}}\gamma_5
(g_t^{\mu\rho}-\frac{1}{3}\gamma_t^{\mu}\gamma_t^{\rho}) ~~~&
{\rm for}~~~ D_{s1}\,,\\[3mm]
\Gamma &=&\displaystyle \sqrt{\frac{1}{2}}
(\gamma_t^{\mu}g_t^{\nu\rho}+\gamma_t^{\nu}g_t^{\mu\rho})
-\frac{1}{3}g_t^{\mu\nu}\gamma_t^{\rho}~~~ &
{\rm for}~~~ D_{s2}^*\,\\[3mm]
\label{current2}
\end{array}
\end{equation}
where $\gamma_t^{\mu}\equiv\gamma^{\mu}-v^{\mu}\!\not\! v$ (In the
rest frame of the meson, $\gamma_t^{\mu}=(0, \vec{\gamma}$)) and
$g_t^{\mu\nu}\equiv g^{\mu\nu}-v^{\mu}v^{\nu}$.  The related decay
constant $f$ is defined as
\begin{equation}
\langle 0|J^{\dag}|D_s^{**}(v, \eta)\rangle \equiv f\eta \,,
\label{decay}
\end{equation}
where $D_s^{**}$ stands for the p-wave meson with polarization
tensor $\eta$, which is the lowest state coupling to $J$.  We will
give the sum rule calculations for masses of the mesons to the
order of $1/m_c$.

\section{QCD Sum Rules}
\label{sec:qcdsr}

The QCD sum rule \cite{qcdsr} is a nonperturbative method rooted in
QCD itself.  We start from the the Green function from which the
hadron mass parameter can be calculated.

\subsection{Green function}
\label{sub:green}

For calculating the masses, the Green function is written as
\begin{equation}
\Pi(\omega)=i\int d^4x e^{ik\cdot x}\langle 0|{\rm T}J^{\dag}(x)J(0)
|0\rangle \,,
\label{pi}
\end{equation}
with $\omega=2k\cdot v$.  It can be expressed in the hadronic language,
\begin{equation}
\Pi(\omega)=\frac{2f^2}{2\bar{\Lambda}_s-\omega}+{\rm higher states}\,.
\label{hadron}
\end{equation}
The calculation of $\Pi(\omega)$ involves perturbation part and
condensation part.  The condensates are included up to dimension
five.  The duality hypothesis is used to simulate the higher
states by perturbative contribution above some threshold
$\omega_c$.  The sum rule is further improved by Borel
transformation. The final sum rule is
\begin{equation}
2f^2e^{-2\bar{\Lambda}_s/T}=\frac{1}{\pi}\int^{\omega_c}_{2m_s}d\nu
{\rm Im} \Pi^{\rm pert}(\nu)e^{-\nu/T}
+\hat{B}^{\omega}_{T}\Pi^{\rm cond}(\omega)\,, \label{sumrulef}
\end{equation}

\subsection{Sum rules in the leading order}
\label{sub:sr}

We found the sum rules in the leading order of the heavy quark
expansion \cite{ours},
\begin{equation}
\begin{array}{lll}
f^2 e^{-2\bar{\Lambda}_s/T}&=&\displaystyle
\frac{3}{64\pi^2}\int_{2m_s}^{\omega_c} (\nu^4+2m_s
\nu^3-6m_s^2 \nu^2\\
&&\displaystyle -12 m_s^3 \nu) e^{-\nu/{T}}d\nu
-\frac{1}{16}\,m_0^2\,\langle\bar ss\rangle
\\[3mm]
&&\displaystyle +{3\over 8}m_s^2
\,\langle\bar ss\rangle-{m_s\over 16\pi}\,\langle \alpha_s G^2
\rangle\;,\\[3mm]
\end{array}
\label{form1}
\end{equation}
for $(0^+,1^+)$ doublet.  And
\begin{equation}
\begin{array}{lll}
f^2 e^{-2\bar\Lambda_s/{T}}&=&\displaystyle\frac{1}{64\pi^2}
\int_{2m_s}^{\omega_c}(\nu^4+2m_s\nu^3-6m_s^2
\nu^2\\
&& \displaystyle -12 m_s^3 \nu) e^{-\nu/{T}}d\nu
-\frac{1}{12}\:m_0^2\:\langle\bar ss\rangle\\[3mm]
&&\displaystyle -\frac{1}{32}\langle\frac{\alpha_s}{\pi}G^2\rangle T
+\frac{1}{8}m_s^2\langle\bar ss\rangle
-\frac{m_s}{48\pi}\langle\frac{\alpha_s}{\pi}G^2\rangle\;,\\[3mm]
\end{array}
\label{form3}
\end{equation}
for $(1^+,2^+)$ doublet. The strange quark mass $m_s$ is kept in
the above calculation. It should be noted that the strange quark
condensate has a different value from up or down quarks.

\subsection{Numerical analysis}
\label{sub:numerical}

In the numerical analysis, we require that the
high-order power corrections be less than $30\%$ of the
perturbation term.  This condition yields the minimum value
$T_{min}$ of the allowed Borel parameter. We also require that the
pole term, which is equal to the sum of the
cut-off perturbative term and the condensation terms, is larger
than $60\%$ of the perturbative term, which leads to the maximum
value $T_{max}$ of the allowed $T$.

We use $m_s = 150 $ MeV for the strange quark mass.  The values of various
QCD condensates are
\begin{eqnarray}
\langle\bar ss\rangle&=&-(0.8\pm 0.1)*(0.24 ~\mbox{GeV})^3\;,\nonumber\\
\langle\alpha_s GG\rangle&=&0.038 ~\mbox{GeV}^4\;,\nonumber\\
m_0^2&=&0.8 ~\mbox{GeV}^2\;.
\label{parameter}
\end{eqnarray}
We use $\Lambda_{QCD}=375$ MeV for three active flavors.

For $(0^+,1^+)$ doublet, the allowed interval of $T$ is determined
to be $0.38 < T < 0.58$ GeV,
\begin{equation}
\bar\Lambda_s(j_l= \frac{1}{2}) = (0.86 \pm 0.10) ~~\mbox{GeV},
\label{result1}
\end{equation}
where the central value corresponds to $T=0.52$ GeV and $\omega_c
=2.9$ GeV.  For $(1^+,2^+)$ doublet, the allowed interval of $T$
is $0.55<T<0.65$ GeV,
\begin{equation}
\bar\Lambda_s(j_l= \frac{3}{2}) = (0.83 \pm 0.10) ~~\mbox{GeV},
\label{result2}
\end{equation}
where the central value corresponds to $T=0.62$ GeV and $\omega_c
=3.0$ GeV.  The errors of the above results refer to the variation
within the stability windows and the uncertainty of $\omega_c$.

\subsection{$1/m_c$ corrections}
\label{sub:1/m}

The $1/m_c$ corrections to $D_s^{**}$ masses come from the Lagrangian of that
order, which is heavy quark symmetry breaking,
\begin{equation}
   {\cal L}_{1/m} = \frac{1}{2 m_Q}\,{\cal K} + \frac{1}{2 m_Q}\,{\cal S} \,,
\label{l1/m}
\end{equation}
with the heavy quark kinetic operator ${\cal K}$ and chromo-magnetic operator
${\cal S}$ being defined as
\begin{equation}
{\cal K}=\bar h_v\,(i D_t)^2 h_v\;,
{\cal S}=\frac{g}{2}\,C_{mag}(m_Q/\mu)\;
   \bar h_v\,\sigma_{\mu\nu} G^{\mu\nu} h_v\;,
\label{ks}
\end{equation}
where $C_{mag}=\displaystyle{\left(\alpha_s(m_Q)
\over\alpha_s(\mu)\right)^{3/{\beta}_0}}$, ${\beta}_0=11-2n_f/3$.
Note that ${\cal S}$ violates both heavy quark flavor and spin symmetries.

The $1/m_Q$ correction to the meson masses is written as
\begin{eqnarray}
\delta M=-{1\over 4m_Q}(K+d_MC_{mag}\Sigma)\;,
\label{delm}
\end{eqnarray}
where $K$ and $d_M\Sigma$ are the mesonic matrix elements of
${\cal K}$ and ${\cal S}$, respectively.  $d_M$ is $(3,-1)$ for
$(0^+,1^+)$ states, and is $(5,-3)$ for $(1^+,2^+)$ states.

In this work, we adopt the method of Ref. \cite{1/m} to use
three-point correlation functions and double Borel transformation.
The $m_s$ terms are neglected at this order. Therefore, the sum
rules are the same as those in \cite{daizhu}. The sum rule window
is taken as the same as that for the leading order. The mixing of
the two $1^+$ states at this order, which was found to be small in
 \cite{daizhu}, is neglected. The numerical results of $K$ and
$\Sigma$ are obtained as follows \cite{ours},
\begin{equation}
\begin{array}{lll}
K(j_l =\frac{1}{2})      & = & (-1.60\pm 0.30) ~~\mbox{GeV}^2\;, \\
\Sigma (j_l=\frac{1}{2}) & = & (0.28\pm 0.05)  ~~\mbox{GeV}^2\;;\\
K(j_l= \frac{3}{2})      & = & (-1.64\pm 0.40) ~~\mbox{GeV}^2\;,\\
\Sigma(j_l=\frac{3}{2})  & = & (0.058\pm 0.01) ~~\mbox{GeV}^2\;.
\end{array}
\label{k}
\end{equation}

\subsection{Final results}
\label{sub:final}

We present our final results in terms of the weighted average mass
and the mass splitting of each doublet,
\begin{equation}
\begin{array}{ll}
{1\over 4}\;(m_{D_{s0}^\ast}+3m_{D^*_{s1}}) &
= m_c+(0.86\pm 0.10)\\
&\displaystyle +{1\over m_c}\;[(0.40\pm 0.08)~\mbox{GeV}^2]\,,
\\[3mm]
m_{D^*_{s1}}-m_{D_{s0}^\ast}&=\displaystyle {1\over m_c}\;[(0.28\pm
0.05)~\mbox{GeV}^2]\;.
\end{array}
\label{f1}
\end{equation}
And
\begin{equation}
\begin{array}{ll}
{1\over 8}\;(3m_{D_{s1}}+5m_{D_{s2}^*})&=
m_c+(0.83\pm 0.10)\\
&\displaystyle +{1\over m_c}\;[(0.41\pm 0.10)~\mbox{GeV}^2]\;,
\\[3mm]
m_{D_{s2}^*}-m_{D_{s1}} & =\displaystyle {1\over m_c}\;[(0.116\pm
0.06)~\mbox{GeV}^2]\;.
\end{array}
\label{f2}
\end{equation}

We obtain the value of $m_c = 1.44$ GeV from the quantity in the
first equation in (\ref{f2}), which is well measured.  This value
is in good agreement with that fitting ground state charm hadrons
\cite{neubert,baryon}.  Then we predict,
\begin{equation}
\begin{array}{ll}
m_{D_{s2}^*}-m_{D_{s1}}=(0.080\pm 0.042)~\mbox{GeV}
&\mbox{Exp 37 MeV},\\
\displaystyle {1\over 4}\;(m_{D_{s0}^\ast}+3m_{D^*_{s1}})
=(2.57\pm 0.12)~\mbox{GeV}&\mbox{Exp 2.42 GeV},\\[3mm]
m_{D^*_{s1}}-m_{D_{s0}^\ast}=(0.19\pm
0.04)~\mbox{GeV}&\mbox{Exp 143 MeV}.
\end{array}
\label{f3}
\end{equation}
The experimental values have been also given for comparison.  The
consistency can be seen.  And
\begin{eqnarray}
m_{D_{s0}^\ast} & = & 2.42 \pm 0.13 ~\mbox{GeV}\;,~~~\mbox{Exp: 2.317 GeV}\,.
\label{f4}
\end{eqnarray}
The experimental values have been also given for comparison.  It
 can be seen that the results of calculations are consistent with the experimental
 data within the uncertainties.
\section{Summary and Discussion}
\label{sec:summary}

The HQET sum rules have been used to the order of $1/m_c$ for
calculating p-wave $D_s$ meson masses.  Within uncertainties of
QCD sum rules, our results for $(0^+,1^+)$ states are consistent
with experimental data for $D_s(2317)$ and $D_s(2457)$. But the
result for the central value of the mass $D_{s0}^\ast$ meson is
still $100$ MeV higher than $D_s(2317)$. The stability of the sum
rules obtained is not as good as that for the ground states. So
the predictions for the masses have large uncertainties, which are
estimated as the errors given. We would like to emphasize that the
result for the mass splitting between these two states agrees with
the experimental value within relatively small uncertainty and
this result is not sensitive to the window taken for the sum rule
for $\Sigma (j_l=\frac{1}{2})$ .

Finally in the same way, we have also calculated the non-strange
p-wave $D$ meson masses \cite{ours}.  The results are consistent
with experimental data within theoretical uncertainties.

The repulsion between the $D_s(0^+)$, $D_s(1^+)$ states and the
$DK$ and $DK^*$ continuum may help to lower their masses. In the
framework of the sum rule this effect should come from the
contribution from the $DK$ and $DK^*$ continuum to the dispersion
integral.

\acknowledgments

The speaker is grateful to Profs. C.W. Kim and Eung Jin Chun for kind
hospitality and support during his stay in KIAS, Seoul.
This project was supported
by the National Natural Science Foundation of China, Ministry of
Education of China and BEPC Opening Project.

\end{document}